\documentclass[a4paper,scale=0.8,onecolumn,nobibnotes,nofootinbib]{revtex4}

\usepackage{amsmath,amssymb}
\usepackage{graphicx,subfigure,epsfig}
\usepackage{hyperref}
\usepackage{geometry}
\hypersetup{
    colorlinks=true,
    citecolor=blue,
    linkcolor=red,
    filecolor=magenta,
    urlcolor=cyan,}

\newcommand{\be}{\begin{equation}}
\newcommand{\ee}{\end{equation}}

\begin{document}

\title{Analysis of the Dirac equation  with the Killingbeck potential in non-commutative space}
\author{Lan Zhong$^{1}$, Hao Chen$^{1}$, Qi-Kang Ran$^{2}$, Chao-Yun Long$^{1}$ and Zheng-Wen Long\footnote{ zwlong@gzu.edu.cn}$^{1}$}
\affiliation{$1$ College of Physics, Guizhou University, Guiyang, 550025, China.\\
 ${2}$ College of Mathematics, Shanghai University of Finance and Economics, Shanghai 200433, China.}

\begin{abstract}
  \textbf{Abstract}. In this paper, we investigate the Dirac equation with the Killingbeck potential under the external magnetic field in non-commutative space. Corresponding to the expressions of the energy level and wave functions in spin symmetry limit and pseudo-spin symmetry limit are derived by using the Bethe ansatz method. The parameter $B$ associated with the external magnetic field and non-commutative parameter $\theta$ make to modify the energy level for considered systems.
\end{abstract}
\maketitle
~~~~~~~~~\textsl{Keywords:} Dirac equation, non-commutative space, Killingbeck potential.

\section{Introduction}	
Killingbeck potential with form $-\frac{1}{r}+2\lambda r+2\lambda^{2} r^{2}$ was introduced by Killingbeck \cite{1} in studying a polynomial perturbation problem about hydrogen atom. The term $-\frac{1}{r}$ represents the gluon exchange potential while the last two terms are for confinement. The Killingbeck potential rewritten as $V(r)=ar^{2}+br-\frac{q}{r}$ has widely studied in various fields of physics, such as atomic and molecular physics \cite{2}, particle physics \cite{3,4}. There are some investigations about the Dirac equation for Killingbeck potential in Refs \cite{5,6}. In the relativistic quantum mechanics, the Dirac equation takes an important part to describe the motion of spin-$\frac{1}{2}$ particles. As one solvable model, there are many ways to study, including: Nikiforov-Uvarov (NU) method \cite{7,8,9}, the asymptotic iteration method (AIM) \cite{10,11} and supersymmetry quantum mechanics (SUSYQM) \cite{12}. If doing momentum transformation, it is well-known as Dirac oscillator \cite{13}, which is a typical system investigated in the curved space \cite{14,15,16,17,18}, and the effect of generalized uncertainty principle on spin-$\frac{1}{2}$ Dirac oscillator and spin-$0$ or $1$ DKP oscillator also are addressed \cite{19,20,21}.\\
Noncommutativity is an interesting topic whatever in mathematics and physics. The first introduction of noncommutative coordinates appeared in the problem of ultraviolet divergence solved by Snyder \cite{22}, this work kept the Lorentz covariance better but it didn't get concerns. With the development about string theory \cite{23,24}, the studies about the low energy effective theory of D-brane with B field showed the noncommutativity among the coordinates on D-brane worldvolume \cite{25,26}.
 In 2000, Shiraz Minwalla et al \cite{27} studied the perturbative dynamics of noncommutative field theories and succeed to explain the appearance of UV/IR mixing in noncommutativity. The effect of the noncommutativity on quantum systems have been widely reported such as the spin Hall effect on noncommutative space \cite{28}, the noncommutative corrections on the persistent charged current \cite{29}, the Landau problem \cite{30} and the HMW effect \cite{31}. The Dirac equation in noncommutative space for hydrogen atom was considered in Ref\cite{32}, this work showed the degeneracy of some energy levels was completely lifted in the NC space. In addition, the Aharonov-Bohm effect for a relativistic spin-half particle have also been analyzed \cite{33,34}. Therefore, it seems interesting to study the Dirac equation with the external magnetic in
non-commutative space by considering the Killingbeck potential.\\
The rest of essays are composed as follow. In the next section, we briefly review the property of non-commutative space. In Section 3, we first derive the general expression of the Dirac equation with scalar and vector potentials and then study the two cases in spin symmetry limit and  pseudo-spin symmetry limit, finally discuss the energy levels for investigated systems. The conclusion is presented in Section 4.
\section{Non-commutative space}
In non-commutative space, the commutative relations between coordinate operator and momentum operator are given by($\hbar=c=1$)
\begin{equation}\label{eq1}
[\hat{x}^{i},\hat{x}^{j}]=i\theta^{ij},[\hat{p}^{i},\hat{p}^{j}]=0,
[\hat{x}^{i},\hat{p}^{j}]=i\delta^{ij}  (i,j=1,2),
\end{equation}
where $\hat{x}^{i}$ and $\hat{p}^{i}$ are coordinate operator and momentum operator in non-commutative space, respectively. $\theta^{ij}=\theta \varepsilon^{ij}$ represents the anti-symmetric matrix, $\theta$ is non-commutative parameter.
The Moyal-Weyl product which is a way to deal with the problem in non-commutative quantum physics reads
\begin{equation}\label{eq2}
(f*g)=\exp(\frac{i}{2}\theta^{ij}\partial_{x^{i}}\partial_{x^{j}})f(x^{i})g(x^{j}),
\end{equation}
here $f(x)$ and $g(x)$ are two arbitrary functions. The Moyal-Weyl product can transform to usual product by Bopp shift\cite{35} and the corresponding expression is
\begin{equation}\label{eq3}
 \hat{x}^{i}=x^{i}-\frac{\theta^{ij}}{2}{p}^{j}, \
 \hat{p}^{i}={p}^{i}.
\end{equation}
with $x^{i}$, ${p}^{i}$ are the coordinate operator and momentum operator in usual quantum mechanics, respectively.\\

\section{The Dirac equation with scalar and vector potentials}
The Dirac equation with the potentials is \cite{36,37}
 \begin{equation}\label{eq4}
[\vec{\alpha} \cdot\vec{p}+\beta (M+S(\hat{r}))]\Psi(\vec{r})=
[E-V(\hat{r})]\Psi(\vec{r}).
\end{equation}
Note that $S(\hat{r})$ is scalar potential while $V(\hat{r})$ is vector potential. $\vec{\alpha}$ and $\beta$
are Dirac matrices
\begin{equation}\label{eq5}
\vec{\alpha_{i}}=\left(\begin{array}{ccc}
0\;\;\;\ \vec{\sigma_{i}}\\
\vec{\sigma_{i}}\;\;\;0\\
\end{array}\right),\end{equation}
\begin{equation} \label{eq6}
\beta=\left(\begin{array}{ccc}
I\;\;\;&0\\
0\;\;\;&-I\\
\end{array}\right).\end{equation}
with $\vec{\sigma_{i}}$ are Pauli's $2\times2$ matrices and I is $2\times2$ unit matrix.
$\Psi(\vec{r})$ as wave function has form
\begin{equation}\label{eq7}
\Psi(\vec{r})=\left(\begin{array}{ccc}
\varphi(\vec{r})\;\;\;&\chi(\vec{r}) \\
\end{array}\right)^{T}.\end{equation}
Combination $Eq.(4)$, $Eq.(5)$, $Eq.(6)$ and $Eq.(7)$, one obtains
\begin{equation}\label{eq8}
\vec{\sigma}\cdot\vec{p}\chi(\vec{r})=(E-M-V(\hat{r})-S(\hat{r}))\varphi(\vec{r}),
\end {equation}
\begin{equation}\label{eq9}
\vec{\sigma}\cdot\vec{p}\varphi(\vec{r})=(E+M-V(\hat{r})+S(\hat{r}))\chi(\vec{r}).
\end {equation}
Motivated by the Ref \cite{38}, taking the exact spin symmetry limit $\Delta(\hat{r})=V(\hat{r})-S(\hat{r})=0$ and the exact pseudo-spin symmetry limit $\Sigma(\hat{r})=V(\hat{r})+S(\hat{r})=0$, we have
\begin{equation}\label{eq10}\begin{array}{l}
\vec{\sigma}\cdot\vec{p}\chi(\vec{r})=(E-M-2V(\hat{r}))\varphi(\vec{r}),\\
\vec{\sigma}\cdot\vec{p}\varphi(\vec{r})=(E+M)\chi(\vec{r})
,\end{array}\end{equation}
\begin{equation}\label{eq11}\begin{array}{l}
\vec{\sigma}\cdot\vec{p}\chi(\vec{r})=(E-M)\varphi(\vec{r}),\\
\vec{\sigma}\cdot\vec{p}\varphi(\vec{r})=(E+M-2V(\hat{r}))\chi(\vec{r}).
\end{array}\end{equation}

\subsection{The case of $\Delta(\hat{r})=0$}
Along the $Z-axis$, the Dirac equation with uniform magnetic filed B is as follow \cite{39}
\begin{equation}\label{eq12}
[(\vec{p}-\frac{e}{c}\vec{A}^{(NC)})^{2}+2(E+M)V(\hat{r})]\varphi(\vec{r})
=[E^{2}-M^{2}]\varphi(\vec{r}).
\end {equation}
$\vec{A}$ is vector potential can be shown $\vec{A}=\vec{A}_{1}+\vec{A}_{2}$,
$\vec{B}=B\hat{Z}$, $\vec{p}$ is replaced by
$\vec{p}\rightarrow \vec{p}-\frac{e}{c}\vec{A}(r)$ with $\vec{p}=-i\hbar\vec{\nabla}$.
The potential in the non-commutative space is given by \cite{40,41,42}
\begin{equation}\label{eq13}
V(\hat{r})=V(r)+\frac{1}{2}(\vec{\theta}\times\vec{p})\cdot\vec{\nabla}V(r)+O(\theta^{2})
=V(r)-\frac{\vec{L}\cdot\vec{\theta}}{2r}\frac{\partial V}{\partial r}+O(\theta^{2}).
\end{equation}
So the vector potential under the additional magnetic flux AB \cite{43} has became
\begin{equation}\label{eq14}
\vec{A_{1}}^{(NC)}=\frac{\vec{B}\times\vec{r}}{2}=\frac{B\hat{r}}{2}\hat{\varphi}
=\frac{B}{2}(r-\frac{L_{z}\theta}{2r})\tilde{\varphi},
\end{equation}
\begin{equation}\label{eq15}
\vec{A_{2}}^{(NC)}=\frac{\Phi_{AB}}{2\pi\hat{r}}\vec{\varphi}
=\frac{\Phi_{AB}}{2\pi}(\frac{1}{r}+\frac{L_{z}\theta}{2r^{3}})\tilde{\varphi},
\end {equation}
\begin{equation}\label{eq16}
\vec{A}^{(NC)}=(\frac{Br}{2}+\frac{\Phi_{AB}}{2\pi r}-\frac{BL_{z}\theta}{4r}+\frac{L_{z}\theta\Phi_{AB}}{4\pi r^{3}})\tilde{\varphi}.
\end{equation}
As same way, in NC space, Killingbeck potential is given by
\begin{equation}\label{eq17}
V(\hat{r})=ar^{2}+br-(q+\frac{L_{z}\theta b}{2})\frac{1}{r}-\frac{L_{z}\theta q}{2}\frac{1}{r^{3}}-aL_{z}\theta+O(\theta^{2}).
\end{equation}
Combination $Eq.(12)$ and $Eq.(16)$, in cylindrical coordinates, there is
\begin{equation}\label{eq18}\begin{array}{l}
\{-[\frac{\partial^{2}}{\partial r^{2}}+\frac{1}{r}\frac{\partial}{\partial r}+\frac{1}{r^{2}}\frac{\partial^{2}}{\partial \varphi^{2}}]+
\frac{e^{2}}{c^{2}}(\frac{Br}{2}+\frac{\Phi_{AB}}{2\pi r}-\frac{BL_{z}\theta}{4r}+\frac{L_{z}\theta\Phi_{AB}}{4\pi r^{3}})^{2}
-\frac{ieBL_{z}\theta}{2cr^{2}}\frac{\partial}{\partial\varphi}\\+\frac{ieB}{c}\frac{\partial}{\partial\varphi}+\frac{ie\Phi_{AB}}{\pi cr^{2}}\frac{\partial}{\partial\varphi}+\frac{ieL_{z}\theta\Phi_{AB}}{2\pi r^{4}}\frac{\partial}{\partial\varphi}+2(E+M)V(\hat{r})+M^{2}-E^{2}\}\varphi(\vec{r})=0.
\end{array}\end{equation}
If we set
\begin{equation}\label{eq19}
\varphi(\vec{r})= e^{(im\varphi)}r^{-\frac{1}{2}}U(r),
\end{equation}
inserting $Eq.(19)$ into $Eq.(18)$, and considering $Eq.(17)$, one obtains
\begin{equation}\label{eq20}\begin{array}{l}
\{\frac{d^{2}}{d r^{2}}+(\frac{1}{4}-m^{2}-\frac{e^{2}\Phi_{AB}^{2}}{4\pi^{2} c^{2}}+\frac{e\Phi_{AB}m}{\pi c}-\frac{eBL_{z}\theta m}{2c})\frac{1}{r^{2}}+(-\frac{e^{2}B^{2}}{4c^{2}}-2(E+M)a)r^{2}\\
-2(E+M)br+(\frac{eL_{z}\theta m\Phi_{AB}}{2\pi c}-\frac{e^{2}L_{z}\theta \Phi_{AB}^{2}}{4c^{2} \pi^{2}})\frac{1}{r^{4}}
+(E+M)L_{z}\theta q\frac{1}{r^{3}}+2(E+M)\\(q+\frac{L_{z}\theta b}{2})\frac{1}{r}
+\frac{e^{2}B^{2}L_{z}\theta}{4c^{2}}-\frac{e^{2}B\Phi_{AB}}{2c^{2} \pi}+\frac{eBm}{c}+2(E+M)aL_{z}\theta
+E^{2}-M^{2}\}U(r)\\=0.
\end{array}\end{equation}
Simplifing $Eq.(20)$ can get
\begin{equation}\label{eq21}
\{\frac{d^{2}}{d r^{2}}+\zeta_{1}\frac{1}{r}+\zeta_{2}\frac{1}{r^{2}}+\zeta_{3}\frac{1}{r^{3}}+\zeta_{4}\frac{1}{r^{4}}
-\zeta_{5}r-\zeta_{6}r^{2}+\zeta_{7}\}U(r)=0,
\end{equation}
with the notations
\begin{equation}\label{eq22}\begin{array}{l}
\zeta_{1}=2(E+M)(q+\frac{L_{z}\theta b}{2}),\\
\zeta_{2}=\frac{1}{4}-m^{2}-\frac{e^{2}\Phi_{AB}^{2}}{4\pi^{2} c^{2}}+\frac{e\Phi_{AB}m}{\pi c}-\frac{eBL_{z}\theta m}{2c},\\
\zeta_{3}=(E+M)L_{z}\theta q,\\
\zeta_{4}=\frac{eL_{z}\theta m\Phi_{AB}}{2\pi c}-\frac{e^{2}L_{z}\theta \Phi_{AB}^{2}}{4c^{2} \pi^{2}},\\
\zeta_{5}=2(E+M)b,\\
\zeta_{6}=\frac{e^{2}B^{2}}{4c^{2}}+2(E+M)a,\\
\zeta_{7}=\frac{e^{2}B^{2}L_{z}\theta}{4c^{2}}-\frac{e^{2}B\Phi_{AB}}{2c^{2} \pi}+\frac{eBm}{c}+2(E+M)aL_{z}\theta
+E^{2}-M^{2}.
\end{array}\end{equation}
Introducing the transformation
 \begin{equation}\label{eq23}
U(r)=\exp({a_{1}r^{2}+a_{2}r+\frac{a_{3}}{r}})\upsilon(r),
\end{equation}
we put $Eq.(23)$ into $Eq.(21)$ and using the Bethe ansatz method\cite{44,45}, we obtain
\begin{equation}
\upsilon(r)=\prod_{i}^{n}\left(r-r_{i}\right), \quad \upsilon(r)=1 \quad for \quad n=1,
\end{equation}
where
\begin{equation}\label{eq25}\begin{array}{l}
a_{3}^{2}+\zeta_{4}=0,\\
4a_{1}a_{2}-\zeta_{5}=0,\\
4a_{1}^{2}-\zeta_{6}=0.
\end{array}\end{equation}
Combination $Eq.(22)$ and $Eq.(25)$, the expression of energy can be got
\begin{equation}\label{eq26}\begin{array}{l}
-2\sqrt{\frac{e^{2}B^{2}}{4c^{2}}+2(E+M)a}(n+\frac{1}{2})+\frac{e^{2}B^{2}L_{z}\theta}{4c^{2}}-\frac{e^{2}B\Phi_{AB}}{2c^{2} \pi}+E^{2}\\-M^{2}+\frac{eBm}{c}+2(E+M)aL_{z}\theta
+\frac{(E+M)^{2}b^{2}}{\frac{e^{2}B^{2}}{4c^{2}}+2(E+M)a}=0.
\end{array}\end{equation}
This is the implicit energy level expression in the case of spin limit with the external magnetic in NC space. In order to analyze the above result, we consider to plot the positive energy eigenvalues $E$ versus the magnetic filed $B$, the non-commutative parameter $\theta$ and potential function parameter $a$, respectively. In the Fig.1, it is about energy eigenvalues and magnetic filed $B$. It directly shows that the energy eigenvalues increase with the increase of the magnetic filed $B$ in the case of different values of quantum member $n$. The Fig.2 and Fig.3 show the variable trend of $E$ via $\theta$ and $a$, respectively. There is the decrease trend of the energy eigenvalues with the large of non-commutative parameter in the Fig.2 while the E has increase trend with a in the Fig.3.

\begin{figure}
\centerline{\includegraphics[width=7.5cm]{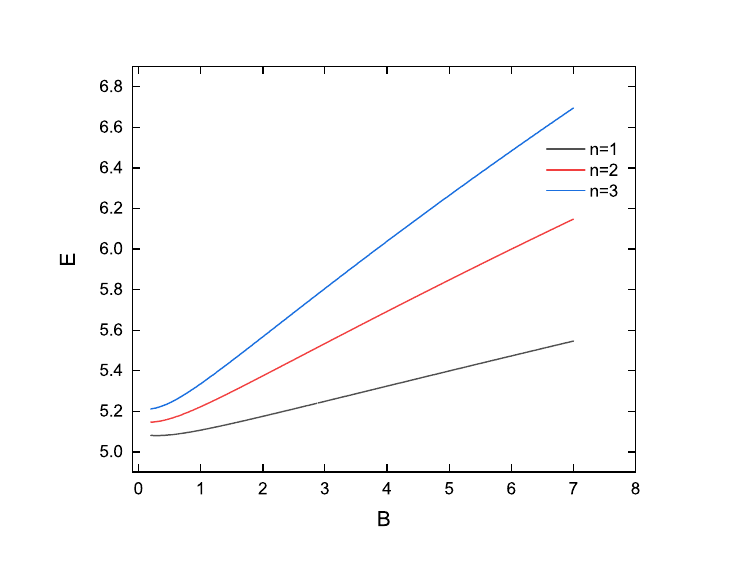}}
\centerline{}
\parbox[c]{15.0cm}{\footnotesize{\bf Fig.~1.}Energy eigenfunctions $E$ for different magnetic filed $B$ with values of the
quantum numbers $n(1,2,3)$,$e=c=L_{z}=m=1,\Phi_{AB}= 2,M=5,a=0.005,b=0.007,\theta=0.001.$}
\end{figure}
\begin{figure}
\centerline{\includegraphics[width=7.5cm]{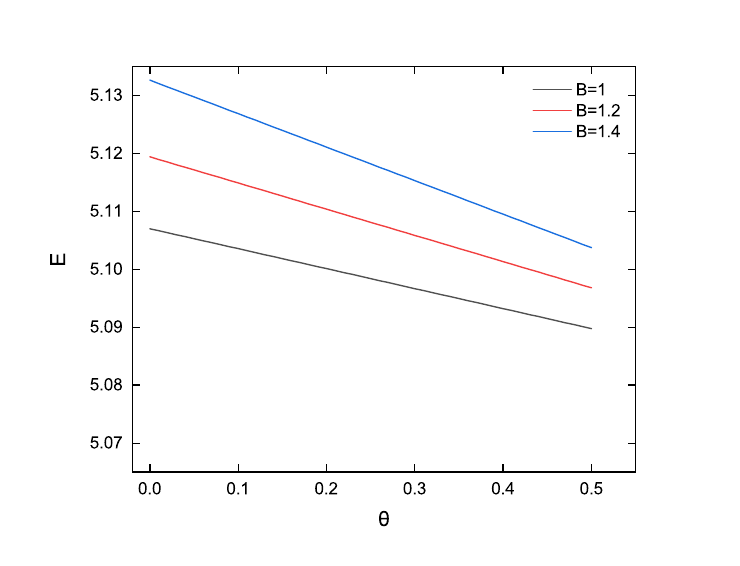}}
\centerline{}
\parbox[c]{15.0cm}{\footnotesize{\bf Fig.~2.}Energy eigenfunctions $E$ for different non-commutative parameter $\theta$ with
values of the magnetic filed $B$(1,1.2,1.4),$e=c=L_{z}=n=m=1,\Phi_{AB}= 2,M=5,a=0.005,b=0.007$ .}
\end{figure}
\begin{figure}
\centerline{\includegraphics[width=7.5cm]{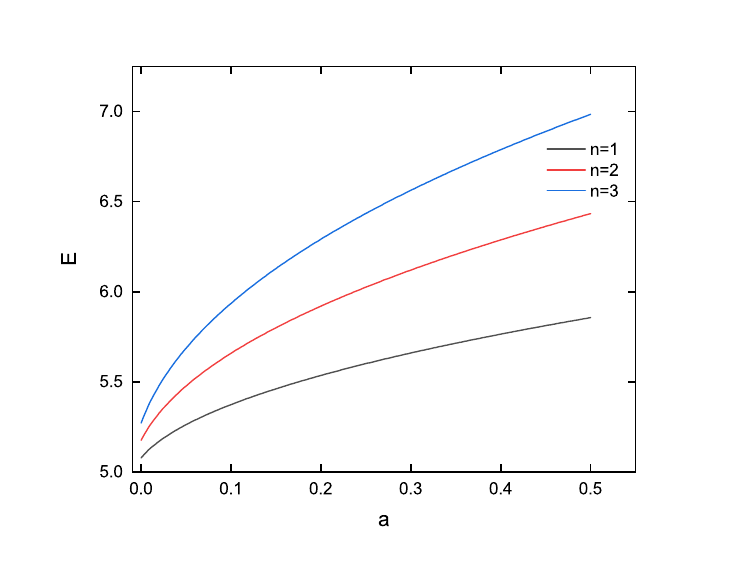}}
\centerline{}
\parbox[c]{15.0cm}{\footnotesize{\bf Fig.~3.}Energy eigenfunctions $E$ for different potential function parameter $a$ with values of the
quantum numbers $n(1,2,3)$,$B=e=c=L_{z}=m=1,\Phi_{AB}= 2,M=5,b=0.007,\theta=0.001$ .}
\end{figure}

And the corresponding wave function is given by
\begin{equation}\label{eq27}
\varphi(\vec{r})= \exp{(im\varphi)}r^{-\frac{1}{2}}\exp({a_{1}r^{2}+a_{2}r+\frac{a_{3}}{r}})\upsilon(r).
\end{equation}
\subsection{The case of $\Sigma(\hat{r})=0$}
In this case, the pseudo-spin symmetry limit $V(\hat{r})=-S(\hat{r})$ is considered. So $Eq.(11)$ under the external magnetic reads
\begin{equation}\label{eq28}
[(\vec{p}-\frac{e}{c}\vec{A}^{NC})^{2}+2(E-M)V(\hat{r})]\chi(\vec{r})
=[E^{2}-M^{2}]\chi(\vec{r}).
\end {equation}
Combination $Eq.(16)$ and the above equation in cylindrical coordinate, we obtain
 \begin{equation}\label{eq29}\begin{array}{l}
\{-[\frac{\partial^{2}}{\partial r^{2}}+\frac{1}{r}\frac{\partial}{\partial r}+\frac{1}{r^{2}}\frac{\partial^{2}}{\partial \varphi^{2}}]+
\frac{e^{2}}{c^{2}}(\frac{Br}{2}+\frac{\Phi_{AB}}{2\pi r}-\frac{BL_{z}\theta}{4r}+\frac{L_{z}\theta\Phi_{AB}}{4\pi r^{3}})^{2}
-\frac{ieBL_{z}\theta}{2cr^{2}}\frac{\partial}{\partial\varphi}\\+\frac{ieB}{c}\frac{\partial}{\partial\varphi}+\frac{ie\Phi_{AB}}{\pi cr^{2}}\frac{\partial}{\partial\varphi}+\frac{ieL_{z}\theta\Phi_{AB}}{2\pi r^{4}}\frac{\partial}{\partial\varphi}+2(E-M)V(\hat{r})+M^{2}-E^{2}\}\chi(\vec{r})=0
.\end{array}\end{equation}
Taking $Eq.(17)$ into $Eq.(29)$ and doing the transformation $\chi(\vec{r})= e^{(il\varphi)}r^{-\frac{1}{2}}\tau(r)$, one obtains
\begin{equation}\label{eq30}\begin{array}{l}
\{\frac{d^{2}}{d r^{2}}+(\frac{1}{4}-l^{2}-\frac{e^{2}\Phi_{AB}^{2}}{4\pi^{2} c^{2}}+\frac{e\Phi_{AB}l}{\pi c}-\frac{eBL_{z}\theta l}{2c})\frac{1}{r^{2}}+(-\frac{e^{2}B^{2}}{4c^{2}}-2(E-M)a)r^{2}\\
-2(E-M)br+(\frac{eL_{z}\theta l\Phi_{AB}}{2\pi c}-\frac{e^{2}L_{z}\theta \Phi_{AB}^{2}}{4c^{2} \pi^{2}})\frac{1}{r^{4}}
+(E-M)L_{z}\theta q\frac{1}{r^{3}}+2(E-M)\\(q+\frac{L_{z}\theta b}{2})\frac{1}{r}
+\frac{e^{2}B^{2}L_{z}\theta}{4c^{2}}-\frac{e^{2}B\Phi_{AB}}{2c^{2} \pi}+\frac{eBl}{c}+2(E-M)aL_{z}\theta
+E^{2}-M^{2}\}\tau(r)\\=0.
\end{array}\end{equation}
The above equation is rewrriten
\begin{equation}\label{eq31}
\{\frac{d^{2}}{d r^{2}}+\xi_{1}\frac{1}{r}+\xi_{2}\frac{1}{r^{2}}+\xi_{3}\frac{1}{r^{3}}+\xi_{4}\frac{1}{r^{4}}
-\xi_{5}r-\xi_{6}r^{2}+\xi_{7}\}\tau(r)=0,
\end{equation}
with the new parameters
\begin{equation}\label{eq32}\begin{array}{l}
\xi_{1}=2(E-M)(q+\frac{L_{z}\theta b}{2}),\\
\xi_{2}=\frac{1}{4}-l^{2}-\frac{e^{2}\Phi_{AB}^{2}}{4\pi^{2} c^{2}}+\frac{e\Phi_{AB}l}{\pi c}-\frac{eBL_{z}\theta l}{2c},\\
\xi_{3}=(E-M)L_{z}\theta q,\\
\xi_{4}=\frac{eL_{z}\theta l\Phi_{AB}}{2\pi c}-\frac{e^{2}L_{z}\theta \Phi_{AB}^{2}}{4c^{2} \pi^{2}},\\
\xi_{5}=2(E-M)b,\\
\xi_{6}=\frac{e^{2}B^{2}}{4c^{2}}+2(E-M)a,\\
\xi_{7}=\frac{e^{2}B^{2}L_{z}\theta}{4c^{2}}-\frac{e^{2}B\Phi_{AB}}{2c^{2} \pi}+\frac{eBm}{c}+2(E-M)aL_{z}\theta
+E^{2}-M^{2}.
\end{array}\end{equation}
We set the auxiliary function
 \begin{equation}\label{eq33}
\tau(r)=\exp({b_{1}r^{2}+b_{2}r+\frac{b_{3}}{r}})\nu(r),
\end{equation}
putting $Eq.(33)$ into $Eq.(31)$ and using the Bethe ansatz method, we obtain
\begin{equation}
\nu(r)=\prod_{i}^{n}\left(r-r_{i}\right), \quad \nu(r)=1 \quad for \quad n=1,
\end{equation}
with
\begin{equation}\label{eq35}\begin{array}{l}
b_{3}^{2}+\xi_{4}=0,\\
4b_{1}b_{2}-\xi_{5}=0,\\
4b_{1}^{2}-\xi_{6}=0.
\end{array}\end{equation}
Combination $Eq.(32)$ and $Eq.(35)$, the implicit expression of E is given by
\begin{equation}\label{eq36}\begin{array}{l}
-2\sqrt{\frac{e^{2}B^{2}}{4c^{2}}+2(E-M)a}(n+\frac{1}{2})+\frac{e^{2}B^{2}L_{z}\theta}{4c^{2}}-\frac{e^{2}B\Phi_{AB}}{2c^{2} \pi}+\frac{eBl}{c}\\+2(E-M)aL_{z}\theta
+E^{2}-M^{2}+\frac{(E-M)^{2}b^{2}}{\frac{e^{2}B^{2}}{4c^{2}}+2(E-M)a}=0.
\end{array}\end{equation}
Take similar ways, we respectively plot the positive energy eigenvalues $E$ versus the magnetic filed $B$, the non-commutative parameters $\theta$ and potential function parameter $a$ under pseudo-spin symmetry. In the Fig.4 and Fig.6, they both show the increase trend of the energy eigenvalues with the large of magnetic filed $B$ and potential function parameter $a$, respectively. In the Fig.5, there is the energy decreases with the non-commutative parameter.
\begin{figure}
\centerline{\includegraphics[width=7.5cm]{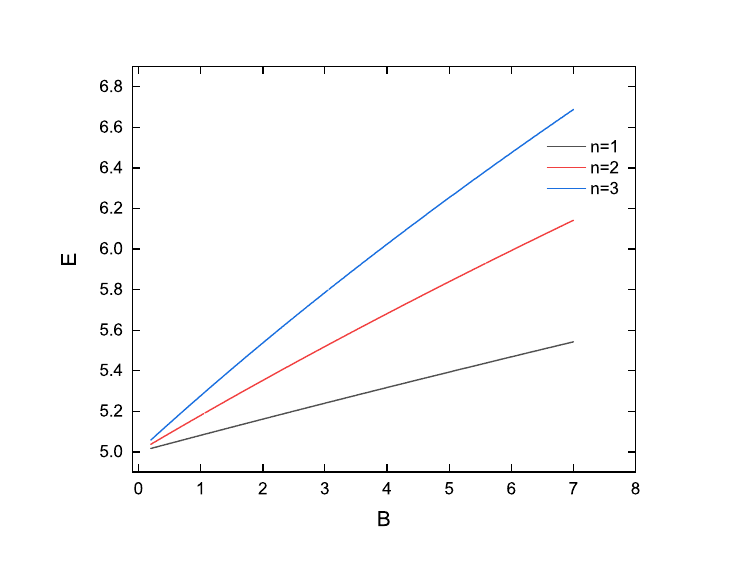}}
\centerline{}
\parbox[c]{15.0cm}{\footnotesize{\bf Fig.~4.}Energy eigenfunctions $E$ for different magnetic filed $B$ with values of the
quantum numbers $n(1,2,3)$,$e=c=L_{z}=l=1,\Phi_{AB}= 2,M=5,a=0.005,b=0.007,\theta=0.001.$ .}
\end{figure}
\begin{figure}
\centerline{\includegraphics[width=7.5cm]{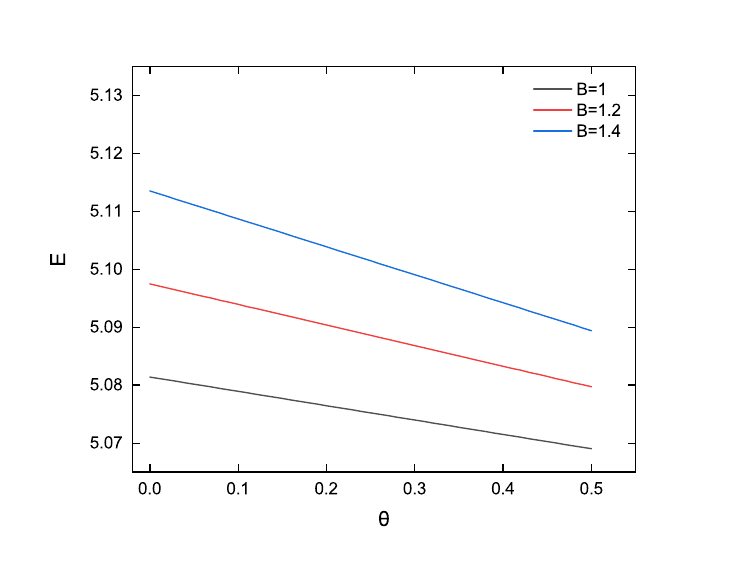}}
\centerline{}
\parbox[c]{15.0cm}{\footnotesize{\bf Fig.~5.}Energy eigenfunctions $E$ for different non-commutative parameter $\theta$ with
values of the magnetic filed $B$(1,1.2,1.4),$e=c=L_{z}=n=l=1,\Phi_{AB}= 2,M=5,a=0.005,b=0.007$ .}
\end{figure}
\begin{figure}
\centerline{\includegraphics[width=7.5cm]{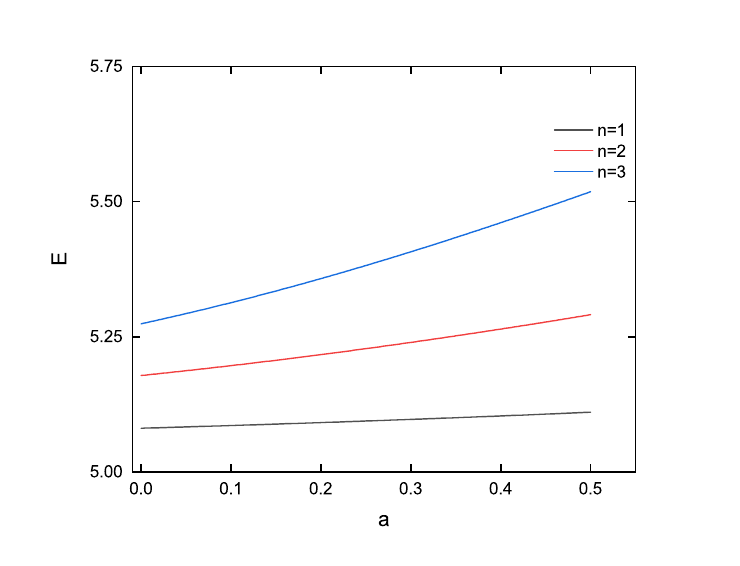}}
\centerline{}
\parbox[c]{15.0cm}{\footnotesize{\bf Fig.~6.}Energy eigenfunctions $E$ for different potential function parameter $a$ with values of the
quantum numbers $n(1,2,3)$,$B=e=c=L_{z}=l=1,\Phi_{AB}= 2,M=5,b=0.007,\theta=0.001$ .}
\end{figure}
And the wave function is given by
\begin{equation}\label{eq37}
\chi(\vec{r})= \exp{(il\varphi)}r^{-\frac{1}{2}}\exp({b_{1}r^{2}+b_{2}r+\frac{b_{3}}{r}})\nu(r).
\end{equation}
\section{Conclusion}
In this work, we have investigated the Dirac equation in the presence of the external magnetic field with the Killingbeck potential in non-commutative space and obtain the implicit expression about energy. Some plots of energy eigenvalues under different magnetic field $B$, potential function parameters $a$ and non-commutative parameters $\theta$ in spin symmetry limit and pseudo-spin symmetry limit have been analyzed in detail. On the one hand, in both cases, the energy eigenvalues respectively show increase trend with the increase of the magnetic field $B$ and potential function parameter $a$ while decrease trend with non-commutative parameter $\theta$, on the other hand there are different values of energy between the case of spin symmetry limit and the case of pseudo-spin symmetry limit. \\

\section*{Acknowledgments}
This work is supported by the National Natural Science Foundation of China (Grant nos. 11465006 and 11565009) and the Major Research Project of innovative Group of Guizhou province (2018-013).


\end{document}